\documentstyle[aps,multicol,epsfig,psfig,pre]{revtex}
\tightenlines
\begin{document}

\title{Unzipping DNA : A hypothesis on correlation during replication
\footnote{Proceedings of CMDAYS2K, held at Guru Ghasidas University,
Bilaspur, India, Aug 29-31, 2K}
}
\author{ Somendra M. Bhattacharjee} 
\address{Institute of Physics, Bhubaneswar 751 005, India\\
  somen@iopb.res.in}
\date{\today}
\maketitle
Based on the result of an unzipping phase transition by a force in a model of 
DNA, we hypothesize that the DnaA-type proteins act as a pulling agent with 
a force
slightly less than the critical force for unzipping.
The dynamic (space-time) correlation of unzipping then drives the
subsequent events of replication.  In such a correlation driven
scenario, there is no need of a replisome as a structural unit 
and this elusive replisome may not exist at all.


\section{Introduction}
\label{sec:Introduction}
A double stranded DNA (dsDNA) needs to be opened up for its
replication\cite{kornberg}. Once done, the subsequent processes take
place sequentially, like the binding of, say, helicases, single-strand-binding
proteins etc, then the elongation of the new chain by polymerases etc.

Recently it has been proposed\cite{smb_mt} that
the action of dnaA-type proteins could be thought of as a force acting at the
initiation site (called ``origin'').  ( The function of DnaA is not to
be confused with the ``locomotive'' or ``sewing machine'' action of
the hypothetical replisome.)  A simple model showed that there is a
phase transition at a critical value of the force so that for forces
less than the critical strength, DNA is a double-stranded object while
for forces exceeding it, the chains get opened up.

In this paper we give a short description of this model, the
principal results and the subsequent developments.  We  point out
the features that seem to have led to ``confusions''.
We also show how this simple model can be extended to
incorporate some other features.  

The connection to in vivo replication and laboratory experiments are
discussed in section III, 
where we propose that the correlation of unzipping is at the heart of the
highly correlated collective process that goes by the name of replication. 

\section{Statistical Mechanical model}
\label{sec:Simple-model}

\subsection{Model and results}
The model proposed originally is that of a pair of flexible polymers
bound together by a short-range interaction and pulled at one end by a 
force\cite{smb_mt}.  We start with a homo-DNA, i.e. all base pairs
identical.  One 
end of the two strands are anchored or tied together and the other end 
is pulled by a force ${\bf g}$.  The Hamiltonian is given by 
\begin{mathletters}
\begin{eqnarray}
  \label{eq:1}
  \frac{H}{k_{\rm B}T}  &=&\int_0^N \! \!\!ds \,
  \left[\frac{\varepsilon_1}{2} \left 
  (\frac{\partial {\bf r}_1(s)}{\partial s}\right
  )^{^{\scriptstyle 2}} + \frac{\varepsilon_2}{2} \left
  (\frac{\partial {\bf r}_2(s)}{\partial s}\right
  )^{^{\scriptstyle 2}} + V({\bf r}(s)) \right ] +
H_{\rm force}, \\
V({\bf r}) &=& v \ \delta({\bf r}),\\
H_{\rm force}&=& -{\bf g}\cdot ({\bf r}(N))=
- {\bf { g}}\cdot \int_0^N \! \!\!ds \, \frac{\partial {\bf
  r}(s)}{\partial s},
\end{eqnarray}
\end{mathletters}
where ${\bf r}_i(s)$ denotes the $d$-dimensional position of a
monomer on chain $i$ (of elastic constant $\varepsilon_i$) at a
contour-length $s$ measured from the tied 
end ($s=0$), $ {\bf r}={\bf r}_1 - {\bf r}_2$ is the relative
separation of the two chains at the same monomer $s$ and $V({\bf r})$
is a short-range potential simulating the 
interaction of the base pairs of two strands.  Any realistic
potential can be chosen here.  But our interest is in the effect of
the force on the bound double-strand case, and so for analytical
simplicity we choose a delta function (or contact)  potential, to be
interpreted as a limit of a  narrow square well.
The strength of the well is chosen to be such that there is only one bound
state in the problem.  

As usual in statistical mechanics, there are two possible ensembles,
namely a constant force (${\bf g}$ constant)  ensemble and a constant
end-point separation(${\bf r}(N)$=constant) ensemble.  For studies of
phase transitions, it is useful to work in 
an ensemble of fixed intensive variable.  We therefore choose the
fixed force ensemble.

For the zero force case,  there is a critical unbinding transition at
$v=v_c$.  The  pulling force would like to align the strands in the
direction of the force while the interaction would like to keep the
strands together.  There is a critical force below which the DNA
remains in the dsDNA phase while for forces exceeding this critical
value it gets unzipped.

Quantitatively, the partition function can be evaluated by a transfer
matrix approach with the contour length as the transfer direction.
The free energy per unit length is given by the largest eigenvalue of
the transfer matrix and a phase transition takes place whenever there
is a degeneracy of the largest eigenvalue.   

The transfer matrix can conveniently be written in the form of
a quantum Hamiltonian, albeit non-hermitian\cite{hatano}, for a
particle of co-ordinate ${\bf r}$ (CM behaviour  is like a 
free chain or a free particle) 
\begin{equation}
\label{eq:qham}
H_q({\bf g}) = \frac{1}{2} ({\bf p} + i {\bf g})^2 + V({\bf r}),
\end{equation}
in units of $\hbar (\equiv k_B T) =1$ and mass $=1$, with ${\bf p}$ as
momentum.  For long chains $N\rightarrow \infty$ the free energy is
the ground-state energy of this non-hermitian Hamiltonian. A phase
transition takes place whenever the ground state is degenerate.  The
analysis done in Ref. \cite{smb_mt} shows that if the ground state
energy ( i.e. the binding energy of the dsDNA per unit length or per
base pair ) is $E_0$, then the critical force is given by
\begin{equation}
  \label{eq:2}
  g_c=2\sqrt{E_0} \sim  |v-v_c|^{1/|2-d|} .
\end{equation}
where the $v$-dependences\cite{kolo} of $E_0$ close to $v_c$, for
general $d$, is used.   
In absence of any force, DNA can be denatured either by changing
temperature or by changing pH of the solution.  In this respect the
effective interaction parameter $v$ is a better variable than the
temperature itself. The generic phase diagram is shown in Fig. 1.

\begin{figure}
\begin{center}
\psfig{file=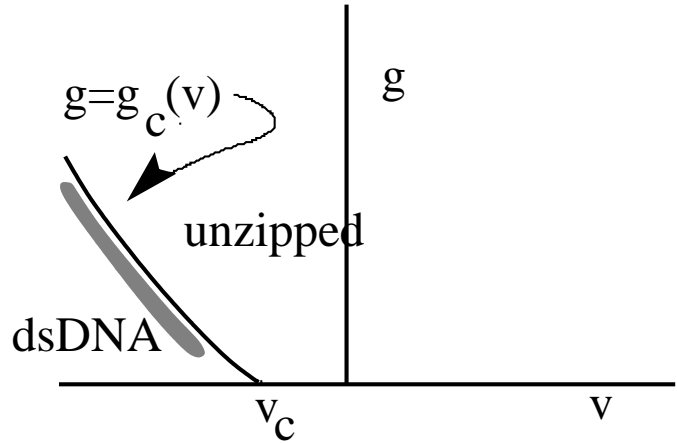,width=3.5in}
\end{center}
\caption{ The $g-v$ phase diagram.  Thermal or pH driven denaturation
without any force is the $g=0$ line.  The DnaA  protein is
hypothesized to work in the shaded region just below the critical line.} 
\label{f.1}
\end{figure}

The degeneracy of the eigenvalues leads to a diverging length scale
that could be associated. for $g \rightarrow g_c -$. as the
characteristic length of the opened Y-fork (the replication fork).
This length scale measured along the strand of the DNA gives a measure
of the number of monomers opened by the force. Denoting this number by
$m$, we get
\begin{equation}
\langle m \rangle \sim \mid g_c-g\mid^{-\nu_{m}} \quad {\rm
    with}\quad \nu_{m}=1.
\label{eq:num}
\end{equation}       
($\nu_m$ was denoted by $\nu_{\tau}$ in Ref \cite{smb_mt}.)

\subsection{Questions on the model and the results}

It is known for centuries that a certain minimum pulling force is
needed to move an object on a surface with friction, or for adhesion,
and so, is the occurrence of a critical force an obvious result?  An
understanding of this threshold phenomena beyond phenomenological
level is still in its infancy.  Our results are to be seen in this
context, and the model is different from the stick-slip models used in
friction or adhesion problems. In fact, previous force
measurements\cite{heslot} for unzipping of DNA have been interpreted
in terms of equilibrium  statistical mechanics,
and not as a friction problem.

A classical diatomic molecule or a bond (or a particle in a potential well)
under a force does not show any
phase transition.  At a quantum mechanical level, a bond with a
stretching force is still described by a hermitian hamiltonian and
therefore will not show the phase transition of the previous subsection.  
It is the polymeric property or connectivity that is crucial for the
unzipping transition.

The result of Eq. \ref{eq:num} is not to be confused with the pulling
of a polymer discussed in text books.  In the case of a polymer pulled
at the ends by a stretching force, there would be an extension of the
chain and for a small force, the overall dimension of the polymer
would scale linearly with the stretching force\cite{degennes}. In
contrast, we see that, in the dsDNA situation, a small force will have
no major effect and the law in Eq. \ref{eq:num} is a dependence on the
{\it deviation from the critical force}.

Is continuum model a good approach?  The answer to this question lies
in the recognition of various length scales.  The Hamiltonian of
Eq. \ref{eq:1} can be discretized and studied with appropriate bond
lengths and realistic $V({\bf r})$.  We adopted a coarse-grained
approach to focus on the phase transition behaviour.  In other words,
a continuum approach by itself is not a necessity to study the phase
transition and other details of unzipping.

Previous studies seem to indicate that DNA is better represented by
worm-like chains but again a renormalization group argument shows that
in the long length limit such a Hamiltonian will generate the elastic
term\cite{freed} and therefore Eq. \ref{eq:1} is a representative of
the universality class.  In the same spirit the delta-function
potential can be replaced by any realistic potential.  But for the
binding-unbinding ( melting) transition, it is only the integral of
the potential that matters and so a square well or delta function is a
valid starting point.

Is the restriction to homo-DNA any good?  It is worth remembering that
the genetic code was deciphered from homoDNA's.  For a controlled
experiment to detect the unzipping transition, large homo-DNA's should
be studied. The model is easily generalizable (see below) to consider
specific base-sequence\cite{rani}.

\subsection{Subsequent developments}
                                       
After the original proposal of model for unzipping in Ref.
\cite{smb_mt}, several studies have been made.  The dynamics of
pulling a polymer from a potential well, i.e., the dynamics of
unzipping in one dimension has been studied at a mean field level by
Sebastian\cite{sebastian}.  In addition to recovering the results of
previous section, Ref \cite{sebastian} shows that the unzipping
transition is also characterized by a diverging time scale. Denoting
the characteristic time scale by ${\sf t}$, one finds
\begin{equation}
\label{eq:time}
{\sf t}\sim \mid g_c-g\mid^{-\nu_{t}} \quad {\rm
    with}\quad \nu_{t}=1.
\end{equation}
A divergence of time scales indicates a long-range correlation in time
in dynamics of unzipping.

DNA generally consists of inhomogeneous sequence of base pairs.  Such
a case in this model can be considered by taking the interaction
energy to be monomer position dependent. The interaction term in
Eq. \ref{eq:1} can be written as $\int_0^{N} \ ds\ b(s) v\ \delta({\bf
r})$, with $b(s)$ a variable depending only on monomer (base) position
$s$ describing the specific details of the sequence.  A case of random
$b(s)$, a random interaction model (RANI model) has been considered
earlier\cite{rani} for the zero force case where the randomness has
been found to be marginal.  In the case of unzipping transition, the
randomness does not destroy the sharpness of the transition but
changes\cite{lubnel} the exponent $\nu_m$ defined in Eq. \ref{eq:num}
to $\nu_m=2$.  The dynamics is yet to be studied.

The nature of the transition has also been considered.  Maritan,
Orlandini and Seno\cite{padova} showed by Monte Carlo simulations and
analytical methods that the noncrossing constraint on the chains can
lead to a first order transition, and more interestingly to a 
``cold denaturation''.  It has recently been
argued\cite{haijun} from a direct evaluation of the partition function
that the transition for the Hamiltonian of Eq. \ref{eq:1} for $d=1$
could be first order.  It is known that depending on the reunion
exponent, the phase transition for polymers could be first
order\cite{fisher84,vsfs,mukamel,grass}, and in the quantum context, 
the order of transition (Eq. \ref{eq:qham}) is
determined by the normalizability of the ``critical''
wavefunction\cite{zia,sixv}.  Several exact results for the unzipping 
transition in various dimensions  have been obtained in Ref.\cite{mtm}.

\subsection{Extensions}

The simple model does not take into account the self avoidance of the
chains.  Effects of self-avoidance has been studied in Ref. 
\cite{padova}.  In a laboratory experiment, one can take Eq. \ref{eq:1} as a
model for DNA in a theta-solvent\cite{degennes}.  In a good solvent,
the self and mutual avoidance of each chain can be introduced by an
imaginary random potential and averaging the partition
function\cite{doi}.  Adding a term to Eq. \ref{eq:1}
\begin{eqnarray}
  \label{eq:imag}
  \frac{H_{\rm imag}}{k_{\rm B}T}  &=&\int_0^N \! \!\!ds \,
 \left [i {\cal V}({\bf r}_1(s)) +  i {\cal V}({\bf r}_2(s)) \right ],
\end{eqnarray}
with ${\cal V}({\bf r})$ as an annealed Gaussian random potential with zero
mean and variance $\langle {\cal V}({\bf r})\ {\cal V}({\bf
r'})\rangle$ = $ u \delta({\bf r}-{\bf r'})$, the averaged partition
function $\overline Z={\overline {\int {\cal D R} \exp(-H)}}$, with
overline denoting averaging over ${\cal V}$, leads to a hamiltonian of
the form of Eq. \ref{eq:1} with additional terms, $H_{\rm int}$
\begin{equation}
\label{eq:saw}
H_{\rm int}=\int ds\ ds' \sum_{i,j=1,2} u \delta({\bf r}_i(s) - {\bf
r}_j(s')).
\end{equation}
This represents excluded volume interaction for the chains. This
imaginary potential of Eq. \ref{eq:imag} would lead to a complex
scalar potential in the quantum hamiltonian of Eq. \ref{eq:qham}.

We have ignored the double helical nature of DNA.  In actual
replication, topoisomerases act when there is super-coiling.  Once the
chains are opened up in a region by DnaA forming a Y-fork, the
remaining chain is brought back to the native sate by the
topoisomerase.  This gives a justification for the absence of the
helical configuration in the hamiltonian in Eq. \ref{eq:1}.  The
topological linking number of the two chains can be described by the
Gauss integral which is known to act like a real vector
potential\cite{doi}.  It would be appropriate to take into account the
linking as an independent quantity only if there is a change in the
coiling in the process\cite{siggia}.

A toy model is a  quantum Hamiltonian corresponding to a polymer with winding 
as 
\begin{equation}
\label{eq:full}
 H_{\rm q} = \frac{1}{2} (p + {\cal A})^2 + V({\bf r}) + i {\cal V}(r),
\end{equation}
where ${\cal A}= A_0({\bf r}) + ig$, with the real vector potential
describing windings and the imaginary part, as before, the exerted
force.  Studies of the most general quantum Hamiltonian with both
complex scalar and vector potential are needed.

\section{Implication for biological replication}

We explain the hypothesis on the role of unzipping in replication.

The replication process starts with a DnaA protein  attaching
itself, in interaction with the membrane, at the ``origin'' to start
the Y-fork ( or ``eye'').  The next step is the binding of the various
enzymes/proteins like single-strand binding (SSB) protein, helicase, 
topoisomerase, polymerases etc.
Unlike the latter proteins/enzymes, the functionality of DnaA protein
is not well understood\cite{plt15}.  It is apparent that the whole
process requires a strong correlation in space and time, whose source
or origin is not obvious.  In order to explain such correlations in
subsequent processes, a structural unit ``replisome'' has been
postulated which has never been isolated.
 
We make the hypothesis that 
\begin{itemize}
\item The function of DnaA is actually to exert a pulling force
(let us take this time $t=0$) with a force which is close to, but
slightly below, the critical force for unzipping.  This force tends to
form a Y-fork or an eye-type bubble (depending on the location of the 
origin).
\item The large length and time scales for the unzipping process,
leads to a space time correlation in the initial nonequilibrium
dynamics of unzipping.  We postulate that this correlation controls
the subsequent processes especially the 
dynamics of bindings of the subsequent enzymes/proteins.
\end{itemize}
The correlation in the unzipping dynamics can be characterized
quantitatively by the early time behaviour of, say, $\langle m(t)
m(t'\rightarrow 0)\rangle$ where $m(t)$ is the number of monomers
unzipped at time $t$, or a more microscopic correlation 
$\langle { r}_i(s,t) { r}_j(s'\rightarrow N,t'\rightarrow 0)\rangle$, $i,j$ 
being the spatial components.

The dynamics of unzipping is described by a Langevin equation
\begin{equation}
  \mu \frac{\partial {\bf r}(s,t)}{\partial t} = - \frac{\delta H}
{\delta{\bf r}(s,t)} + \eta(s,t),
\end{equation}
where $t$ represents time, $H$ is given by Eq. ~\ref{eq:1}, $\eta$ is
the thermal noise related to friction coefficient $\mu$ by
$\langle\eta_i(s,t)\eta_j(s',t')\rangle=2\delta_{ij} \mu k_BT\delta(s-s')\ 
\delta(t
-t')$.  The binding of the next set of proteins in the opened region
may be represented symbolically by a kinetic equation of the type
\begin{equation}
\frac{\partial c}{\partial t}=\ -  {\cal C}(s,t)\  {\sf f}\{c\},
\end{equation}
where $c$ denotes the concentration of the free enzymes in the bath,
${\sf f}\{c\}$ describes the dynamics in absence of any correlations,
and ${\cal C}(s,t)$ stands for the initial or early time behaviour of
the unzipping correlations.  This is equivalent to making the rate
constant dependent on the initial correlation\cite{zwanzig}.
This way of viewing the replication process as coupled system driven
by the long-range space-time correlated unzipping dynamics is different 
from the prevalent contact-based approach.

It is necessary to {\it (a)} study long homo-DNA's under well-controlled 
pulling force to map out the phase diagram \cite{smb_mt,padova}, and 
{\it (b)} probe jointly the correlation and binding kinetics in presence 
of a pulling force.

A few appealing features of such a scenario are to be noted. Since the
critical force is dependent on the structural details of the DNA,
which is different for different species, our hypothesis naturally
requires species variations of the DnaA-type protein for proper
functionality.  This protein should be such that it exerts a force
slightly below the critical value for that particular DNA.  Finally,
in the proposed scenario, the dynamic correlation plays the most
important role, and therefore a composite structure is not a necessity.
In other words, the so-far-elusive hypothesized replisome 
need not exist as a structural unit.



\begin{references}
\bibitem{kornberg}  A. Kornberg  and T.  Baker,   {\it DNA Replication},
  (3rd ed., W. H. Freeman, 1992).  
\bibitem{smb_mt} S. M. Bhattacharjee, {\it Unzipping DNA: towards the
first step of replication}, cond-mat/9912297 (available from 
http://xxx.lanl.gov or its mirror). J. Phys. A {\bf 33}, L423 (2000)
\bibitem{hatano} N. Hatano and  D. R.  Nelson, 
Phys. Rev. B {\bf 56} 8651 (1997).
\bibitem{kolo} E. B. Kolomeisky and J. P. Straley, Phys. Rev. B {\bf
    46}, 12664 (1992).
\bibitem{heslot} U. Bockelmann, B.  Essevaz-Roulet and F. Heslot,
  Phys. Rev. E {\bf 58} (1998) 2386. 
\bibitem{degennes} P. G. de Gennes, {\it Scaling concepts in polymer
physics}, Cornell University Press, Ithaca (1979).
\bibitem{freed} K. Freed, {\it Renormalization group theory of
    macromolecules}, John Wiley, NY (1987).
\bibitem{sebastian} K. L. Sebastian, Phys. Rev. E {\bf 62}, 1128 (2000). 
\bibitem{rani} S. M. Bhattacharjee and S. Mukherji,
  Phys. Rev. Lett. {\bf 70} 49 (1993); {\bf 70}, 3359(E) (1993); 
  Phys. Rev. E {\bf 48},  3483 (1993).
\bibitem{lubnel} D. Lubensky and D. R. Nelson, Phys. Rev. Letts. {\bf 85},
     1572 (2000).
\bibitem{padova} A. Maritan, Orlandini, F. Seno, University of
Padova preprint (2000)
\bibitem{haijun} Haijun Zhou, cond-mat/0007015 (available from 
http://xxx.lanl.gov or its mirror)
\bibitem{fisher84} M. E. Fisher, J. Stat. Phys.  {\bf 34}, 667 (1984).
\bibitem{vsfs} S. M. Bhattacharjee and S. Mukherji, Phys. Rev. Lett.
 {\bf 83}, 2374 (1999).
\bibitem{mukamel} Y. Kafri, D. Mukamel and L. Peliti,
  cond-mat/0007141 (available from http://xxx.lanl.gov or its mirror).
\bibitem{grass} M. S. Causo,B. Coluzzi and P. Grassberger,
  Phys. Rev. E {\bf 62}, 3958 (2000). 
\bibitem{zia}  R. K. P. Zia,  R. Lipowsky and D. M. Kroll,  Am. J. Phys 
{\bf 56}, 160 (1988); S. Mukherji and S. M. Bhattacharjee, under preparation
\bibitem{sixv} Another well-known example of criticality with first-order 
transition is the ferroelectric six vertex model. J. F. Nagle, Comm. Math. 
Phys. {\bf 13}, 62 (1969); R. J. Baxter, {\it Exactly solved models in 
statistical physics}, Academic Press,  1982. 
\bibitem{mtm} D. Marenduzzo, A. Trovato and A. Maritan, preprint (2000). 
\bibitem{doi} S. F. Edwards,  in {\it Polymer physics: 25 years of the
Edwards Hamiltonian}, Ed. by S. M. Bhattacharjee, World Scientific,
Singapore (1992); M. Doi and S. F. Edwards, {\it Dynamics of polymer
solutions}, Oxford U. Press (1986). 
\bibitem{siggia} J. F. Marko and E. D. Siggia, Phys. Rev. E {\bf 52}, 2912 
(1995). 
\bibitem{zwanzig} R. Zwanzig, J. Chem. Phys.  {\bf 97}, 3587 (1992).
\bibitem{plt15} See e.g. Plate 15 (NOT Chapter 15) of Ref 1.
\end{references}
\end{document}